\documentclass[aps,12pt,superscriptaddress,preprint]{revtex4-1}
\usepackage{graphicx}
\usepackage{amsmath}
\usepackage{hyperref}
\usepackage{bm}
\usepackage[dvips]{color}

\newcommand{\fzd} {Helmholtz-Zentrum Dresden-Rossendorf, Institute of Ion Beam Physics and Materials Research, P.O. Box 510119, 01314 Dresden, Germany}

\newcommand{\tud} {Technische Universit\"{a}t Dresden, 01062 Dresden, Germany}

\begin{document}

\title{Hyperdoping Si with chalcogen: solid vs. liquid phase epitaxy}
\date{\today}

\author{Shengqiang~Zhou}
\email[Electronic address: ]{s.zhou@hzdr.de}
\address{\fzd}
\author{Fang~Liu}
\address{\fzd}
\address{\tud}
\author{S.~Prucnal}
\address{\fzd}
\author{Kun Gao}
\address{\fzd}
\address{\tud}
\author{M.~Khalid}
\author{C.~Baehtz}
\author{M.~Posselt}
\author{W.~Skorupa}
\address{\fzd}
\author{M.~Helm}
\address{\fzd}
\address{\tud}

\begin{abstract}
Chalcogen-hyperdoped silicon shows potential applications in
silicon-based infrared photodetectors and intermediate band solar
cells. Due to the low solid solubility limits of chalcogen
elements in silicon, these materials were previously realized by
femtosecond or nanosecond laser annealing of implanted silicon or
bare silicon in certain background gases. The high energy density
deposited on the silicon surface leads to a liquid phase and the
fast recrystallization velocity allows trapping of chalcogen into
the silicon matrix. However, this method encounters the problem of
surface segregation. In this paper, we propose a solid phase
processing by flash-lamp annealing in the millisecond range, which
is in between the conventional rapid thermal annealing and pulsed
laser annealing. Flash lamp annealed selenium-implanted silicon
shows a substitutional fraction of $\sim$ 70\% with an implanted
concentration up to 2.3\%. The resistivity is lower and the
carrier mobility is higher than those of nanosecond pulsed laser
annealed samples. Our results show that flash-lamp annealing is
superior to laser annealing in preventing surface segregation and
in allowing scalability.
\end{abstract}

\section*{Introduction}
Chalcogen-hyperdoped silicon (Si) much above the solid solubility
limits has been investigated due to its specific physical
properties, such as a near-unity broadband (particularly below the
Si bandgap) absorption
\cite{Kim2006APL,BobJAP2010,Tabbal2010,Umezu2013JAP}, a large
enhancement of below-band-gap photocurrent generation
\cite{PSSA201228202}, and insulator-to-metal transition
\cite{PhysRevLett.106.178701,PhysRevLett.108.026401}. These novel
properties make chalcogen-hyperdoped Si a promising material for
applications in Si-based infrared photodetectors
\cite{Carey:05,SaidAPL2011,0268-1242-27-10-102002} and
intermediate band solar cells
\cite{PhysRevLett.78.5014,Antolin2009APL,PhysRevB.82.165201,guenther2014}.
The insulator-to-metal transition realized in sulfur and selenium doped Si is driven by deep level
impurities, which opens a new material test-bed for insulator-to-metal transition, the
long-standing fundamental problem. Note that chalcogen elements in
Si generally have much lower solubility limits compared with III
and V-column elements. For instance, the solid solubility limit
for sulfur in Si \cite{Carlson195981} was determined to be around
$10^{16}$ cm$^{-3}$. To realize hyperdoping above the solid
solubility limits, non-equilibrium methods were used. Femtosecond
(fs) or nanosecond (ns) pulsed laser irradiation was successfully
applied to prepare chalcogen-hyperdoped Si by melting the Si
surface in certain background gases containing chalcogens or with
a powder/film of chalcogen on the surface
\cite{sheehy2007chalcogen,Tabbal2010,smith2013improving}. In
parallel, by ion implantation and subsequent annealing with ns
laser pulses chalcogen-hyperdoped Si have been fabricated
\cite{Kim2006APL,tabbalJVST2007,BobJAP2010,Umezu2013JAP}. Both fs
and ns (or even longer pulse length) laser annealing (pulsed laser annealing: PLA) at a high
enough energy density melts the Si surface \cite{stritzker1981measurement} and renders a fast
solidification after the laser is off. The solute impurities are
trapped at the moving interface when the regrowth time is short
enough
\cite{CampisanoAPL1980,aziz1982model,aziz1986solute,aziz1988continuous}.
However, the liquid phase epitaxy associated with fs or ns laser
annealing encounters some drawbacks. One is the dopant
redistribution and pileup at the surface
\cite{celler1978spatially,CampisanoAPL1980,BobJAP2010}. The second
is the significant loss of the dopant due to surface evaporation
\cite{BobJAP2010}. The third is the formation of polycrystalline
materials mainly due to the rejected excess dopants
\cite{CampisanoAPL1980,dubon2006doping}. The last but not the
least is its missing scalability and reproducibility of PLA, which
inhibits the application in microelectronics industry.

In this paper, we propose a novel approach, flash lamp annealing
(FLA) in millisecond time duration which allows for a solid phase
epitaxy from implanted Si layers, to realize selenium hyperdoped
Si. We will show that the solid phase epitaxy also can achieve
selenium-hyperdoped Si with larger Hall mobility. Additionally,
the problems mentioned above occurring in nanosecond pulsed laser annealing
can be solved by flashlamp annealing.

\section*{Results}

\subsection{Selenium substitution without diffusion}
Figure 1 shows the representative Rutherford
backscattering spectrometry (RBS) random and channeling spectra of
selenium-implanted Si substrates (Sample SiSe2.3 in table
\ref{tab:sample}) that were pulsed laser annealed (wavelength: 308 nm, pulse length: 28 ns) or flash lamp
annealed (pulse length: 1.3--20 millisecond). We also measured the random and
channeling spectra of the unimplanted virgin substrate and the
channeling spectrum for the as-implanted wafers (not shown to
avoid overlapping). The implanted layer is fully amorphous. As shown in Fig. 1(a), the implanted layer can be well recrystallized by both annealing methods. The RBS-channeling spectrum reveals a near-surface minimum backscattered yield $\chi_{min}$ (the ratio of the aligned to random yields) of about 5\%, which is very close to the value of 4\% determined for the virgin single crystal Si substrate.

\begin{figure}
\includegraphics[scale=0.4]{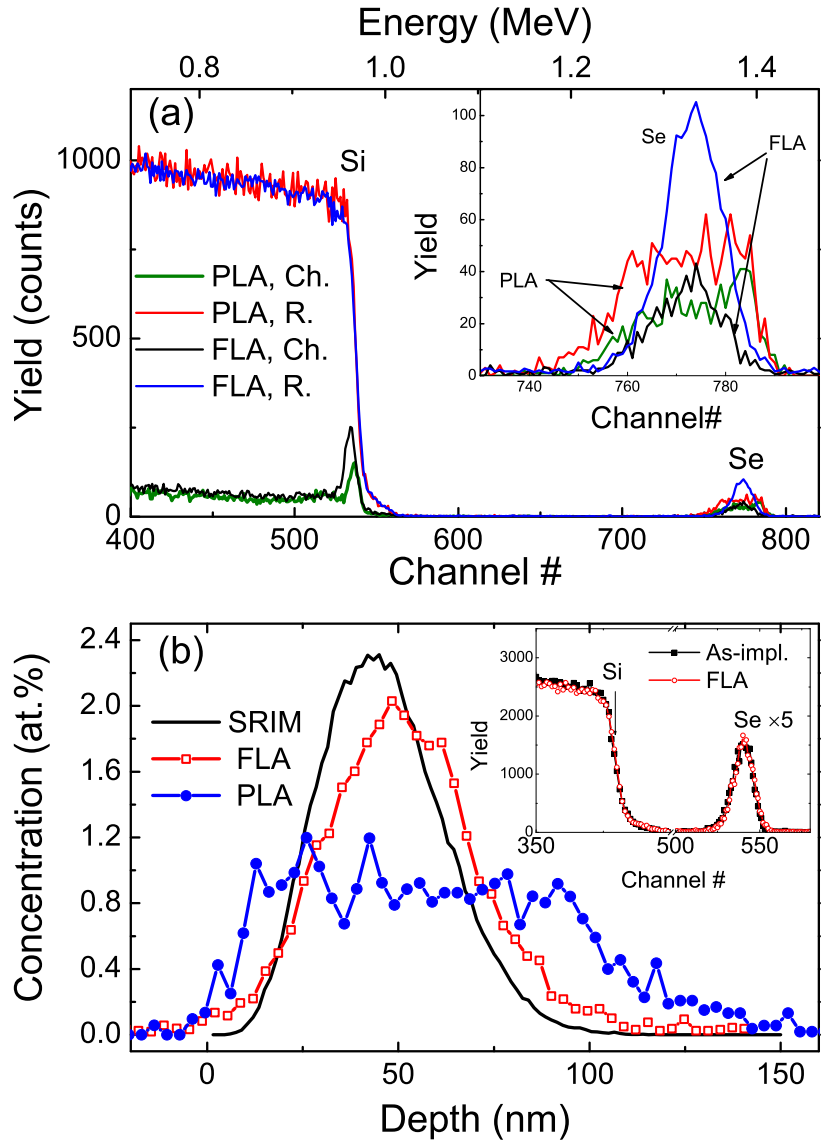}
\caption{(a) A sequence of 1.7 MeV He RBS/channeling spectra of
selenium implanted Si single crystals after different annealing
(taking sample SiSe2.3 as the example). The channeling is along
Si[001]. The Si matrix is recrystallized after both FLA (1.3 ms, 3.4 kV) and PLA (308 nm, 28 ns, 0.9 J/cm$^2$). The inset shows a zoom for the selenium signal. We can see that selenium ions are mostly substitutional to the Si sites. However, for the PLA sample, selenium ions do not
substitute Si sites at the near surface range. (b) The depth profile of selenium in Si after FLA or PLA calculated from RBS spectra. The projected range of selenium in FLA samples is in a reasonable agreement with SRIM simulation after considering the well-known discrepancy in the projection range between SRIM simulation and the experimental values \cite{fichtner1987range,palmetshofer2001range,jin2013ion}. After PLA, a significant re-distribution of selenium is observed. The inset shows the RBS spectra for the as-implanted and FLA samples measured using 1.5 MeV He ions. There is no re-distribution of selenium after FLA.}\label{fig_RBS_PLAFLA}
\end{figure}

In the selenium signal, two interesting features can be observed. One is
the channeling effect, which occurs in both FLA and PLA samples.
It proves the substitution of selenium onto the Si lattice sites, which
is expected and has also been proved by other experiments
\cite{CampisanoAPL1980}. The substitutional fraction can be
approximated by 1-$\chi_{min}$(selenium), which is shown in Fig. 2 for different samples. $\chi_{min}$(selenium) stands for the ratio between the channeling and the random spectra for
the selenium signal. As shown in Fig. 2(a) and
(c), the selenium substitutional fraction is higher in the FLA sample
than in the PLA sample. Another feature is the significant
redistribution of selenium upon PLA. As shown in Fig. 1(b), selenium atoms move both inward and outward upon PLA, which has been reported in many papers
\cite{CampisanoAPL1980,PhysRevLett.41.1246,BobJAP2010,Recht78726}.
However, there is no observable redistribution in the FLA sample as shown in the inset to Fig. 1(b). In RBS spectra, the selenium signal of the FLA sample almost overlaps
with that of the as-implanted sample. The so-called snow plough effect \cite{baeri2008dependence,PhysRevLett.41.1246} is avoided in solid phase epitaxy. In Fig. 1(b), we also show the depth distribution selenium obtained by SRIM (Stopping and Range of Ions in Matter) simulation \cite{ziegler2008srim}. The slightly shallow and narrow distribution of selenium from simulation is well known \cite{fichtner1987range,palmetshofer2001range} and is possibly due to the overestimation of electronic stopping for heavy ions in semiconductors \cite{jin2013ion}.

In Figure 2 we compare the selenium
substitutional fraction for samples with different selenium
concentration and with different annealing parameters. Figure 2(a) shows the substitutional fraction after FLA at different temperatures with the pulse duration of 1.3
ms. The temperature is estimated according to calibration and theory model \cite{skorupa2005advanced}. All the samples are recrystallized. The substitutional
fraction is between 40 and 70\% and shows a maximum at the
annealing temperature of 1473 K. The substitutional fraction
decreases with increasing the flash lamp pulse duration from 1.3
ms to 20 ms as shown in Fig. 2(b). This can be expected, since the diffusion length will be much larger if
the sample is kept at higher temperature for longer time. For PLA
samples, when the laser energy density is below 0.6 J/cm$^2$ the
sample is not fully crystallized according to our RBS/channeling measurements, particulary for the sample with
smaller selenium concentration. The laser annealing parameters have been
found to sensitively depend on the implantation fluence
\cite{venkatesan1978dose,bean1979substrate}. As shown in Fig. 2(c), when the pulsed laser annealing energy is 0.5 J/cm$^2$, the substitutional fraction in sample SiSe1.1 is nearly
zero. When the energy is high enough (larger than 0.6 J/cm$^2$ for
both samples) to recrystallize the full layer, there is not much
difference in the substitutional fraction. Yet the substitutional
fraction for PLA samples is generally lower than for FLA samples.
It is worthy to note that for both FLA or PLA samples, the
substitutional fraction for sample SiSe1.1 is slightly smaller
than for sample SiSe2.3, for which the reason is not clear and more investigation is required. The substitutional fraction has been estimated by comparing the integration of the selenium signal in channeling and random RBS spectra. Therefore, the relative error in the substitutional fraction values when comparing different samples is very small.

\begin{figure}
\includegraphics[scale=0.4]{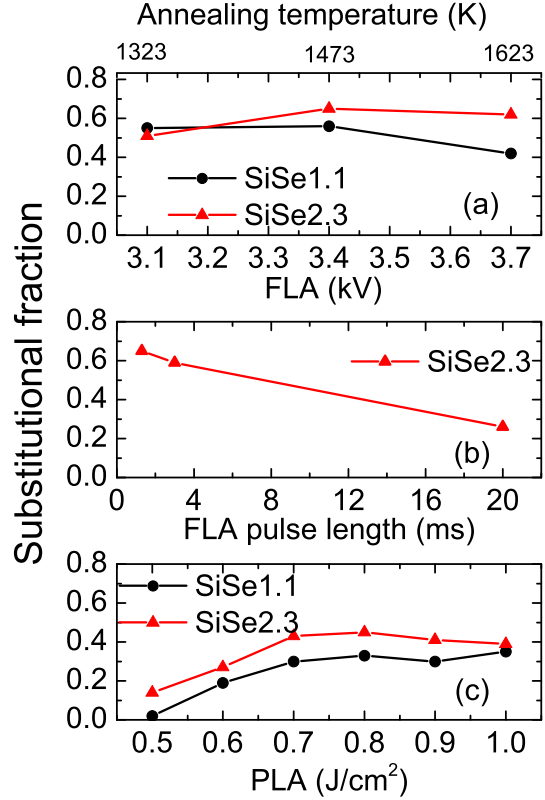}
\caption{The selenium substitutional fraction in selenium implanted Si after
annealing by flash lamp or pulsed laser at different annealing
energy: (a) FLA at different flash-lamp charging voltages
(different annealing temperatures) with the pulse duration of 1.3
ms, (b) FLA with different pulse duration and different charging voltages to have similar peak temperature ($\sim$ 1473 K) in the sample and (c) PLA (308 nm, 28 ns) with different
energy densities.}\label{Fig_Substutionalrate}
\end{figure}

\subsection{Structural properties: epitaxial recrystallization}
The structural properties were characterized by Raman scattering
and by X-ray diffraction (XRD). Figure 3(a) shows
the Raman spectra obtained from selenium implanted Si as well as a
reference single crystalline Si. The as-implanted sample shows a
broad Raman band at around 460 cm$^{-1}$, which corresponds to the
amorphous silicon formed during ion implantation
\cite{prucnal2010formation}. After PLA or FLA, both samples
exhibit a peak at 520 cm$^{-1}$, corresponding to the transverse
optical (TO) phonon mode of crystalline Si. Moreover, an
additional peak at about 300 cm$^{-1}$ which corresponds to the
second-order transverse-acoustic phonon (2TA) scattering from
crystalline Si is observed for both samples as for the Si
reference sample, which is a fingerprint of high crystallinity of
the probed layer. The Raman results prove the complete recrystallization of selenium implanted Si by both PLA and FLA
with relatively high quality.

\begin{figure}
\includegraphics[scale=0.4]{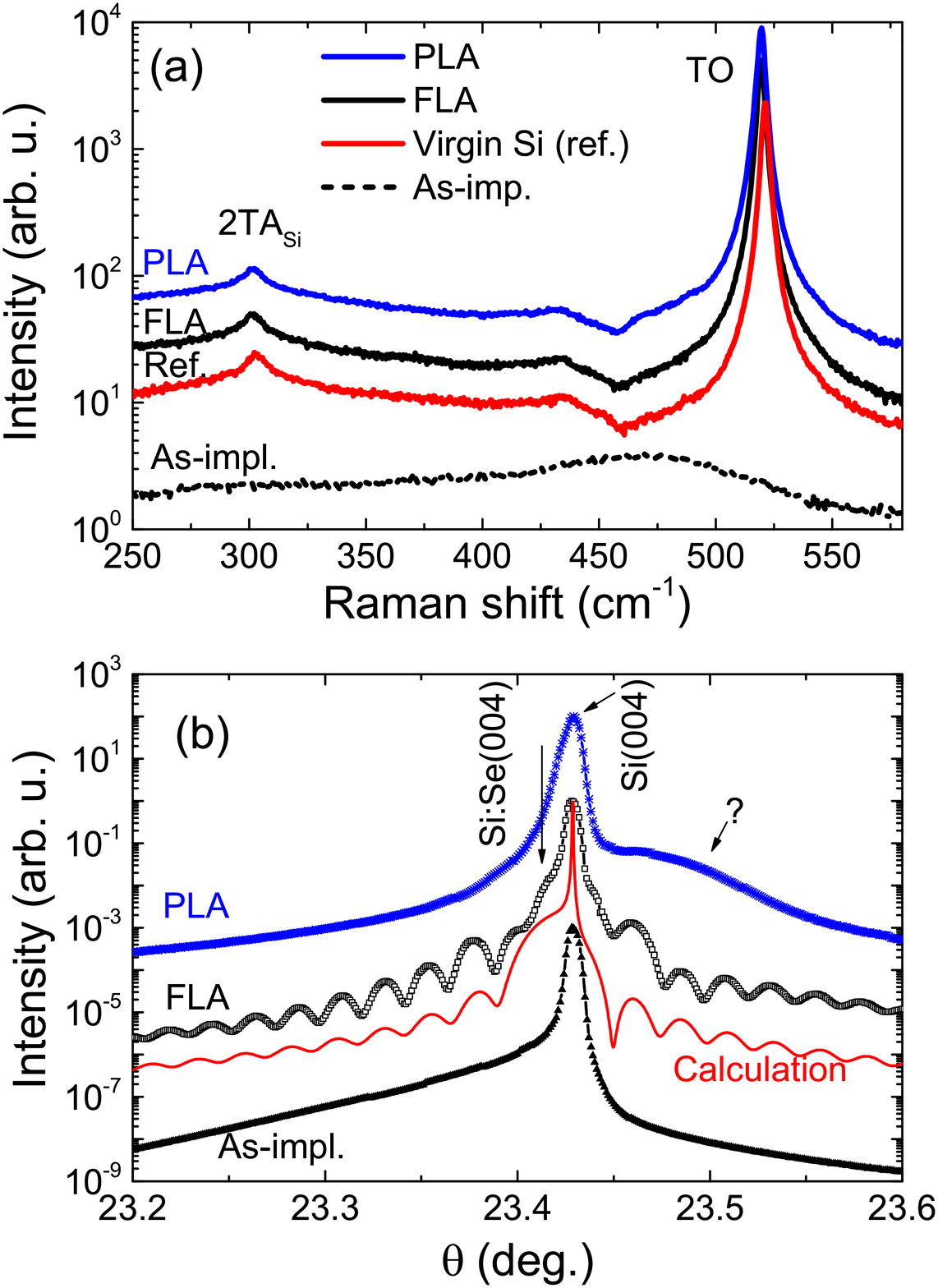}
\caption{Representative Raman or XRD results taking SiSe0.9 as the
example after FLA (1.3 ms, 3.4 kV) and PLA (308 nm, 28 ns, 0.9 J/cm$^2$) (a) $\mu$-Raman spectra: A virgin Si is also shown for comparison. The spectra have been vertically offset for clarity. (b) X-ray diffraction
$\theta-2\theta$ scans: The broad shoulder at the left side for the
as-implanted sample is due to the ion beam induced damage. After
annealing, this broad shoulder disappears and the oscillation
indicates the good crystalline quality.}\label{fig_RamanXRD}
\end{figure}

As shown in Fig. 3(b), in the XRD
$\theta-2\theta$ scans the as-implanted sample show a broad
shoulder at the left side (lower angles). It is due to the
crystalline damage induced by ion implantation. The implanted
layer is amorphous-like as confirmed by Raman scattering (Fig. 3(a)). After annealing, the left-side shoulder disappears, indicating the recrystallization. There is large
difference in the XRD patterns between the PLA and the FLA
samples. When selenium ions substitute the Si sites, they form covalent
bonds. The covalent radii for selenium and Si are 120 and 111 pm,
respectively. The doping of selenium in the Si matrix up to a large
enough concentration is expected to increase the lattice
parameter, like in antimony doped Si
\cite{radamson1994characterization}. We do observe this effect in
the FLA sample. Note that the appearance of the x-ray interference
effect (Pendell\"{o}sung) in the FLA sample is due to two layers
with different lattice spacing. By fitting the XRD pattern, we
obtain a top Si:Se layer with a lattice constant larger by around
0.60\% compared with the Si substrate. The presence of the
interference peaks also proves the high crystalline quality and
the sharp interface after flashlamp annealing. However, the XRD
result for the PLA sample is rather puzzling: it contains a
broader shoulder at a larger diffraction angle corresponding to a
smaller lattice constant.

\subsection{Electrical properties: selenium hyperdoping}

Selenium is a deep donor in Si and its energy level is around 200--300
meV below the Si conduction band \cite{pantelides1992deep}. Upon
high concentration doping, an insulator-metal transition was
observed in selenium doped Si \cite{PhysRevLett.108.026401}. We also
measured the electrical properties of selected samples. Figure 4 shows the sheet resistance in the temperature range 4–-30 K. Since we use nearly intrinsic Si substrate with a sheet resistance above 10$^7$ ohm at room temperature, the parallel resistance from the substrate is much larger than the selenium doped layer. Therefore, we only measure the conductivity from the doped Si layer. For the PLA samples, an insulator-metal transition occurs with increasing selenium concentration: Sample SiSe1.1 behaviors like an insulator with its sheet resistance sharply rising at low temperature. Its conductivity is thermally activated. On the other hand, for the higher doped  sample SiSe2.3 the resistance increases only very slightly at low temperature and its conductivity appears to remain finite when the temperature approaches zero. In sharp contrast, flash lamp annealing renders both samples metallic like -– the higher doped SiSe2.3, but also the lower doped SiSe1.1. The sheet resistance of sample SiSe1.1FLA is even lower than SiSe2.3PLA with a higher Se concentration, clearly showing the superior (flash-lamp) annealing behavior by solid-phase epitaxy. Finally sample SiSe2.3FLA exhibits the smallest sheet resistance and a clear metal-like conductivity. Its sheet resistance at 5 K is around 190 ohm/square. It corresponds to a conductivity of 500/(ohm$\cdot$cm)
if assuming a thickness of 100 nm. This conductivity is three
times larger than sample SiSe2.3PLA. We attribute the large
conductivity to the high quality of the recrystallized layer by
FLA, which results in a large Hall mobility.

Figure 5 shows the comparison of the electron
concentration and Hall mobilities measured at 300 K for different
FLA and PLA samples with increasing selenium concentration. The carrier
concentration is calculated by assuming the effective thickness of
the selenium doped layer as 150 nm for sample SiSe0.9 and 100 nm for
samples SiSe1.1 and SiSe2.3 (see Fig. 1),
respectively. As can be seen, the carrier concentration is in the
range of 3--13$\times10^{19}$ cm$^{-3}$. The activation efficiency
is thus around 10\%. Although the PLA samples generally have a
larger electron concentration than the FLA samples, the mobility
is around 3 times smaller. The Hall mobility in FLA samples is in
the range of 80--100 cm$^{2}$/$V{\cdot}s$. The values achieved
here are almost of the same order as in Si doped with normal
shallow donors in the high doping regime (10$^{19-20}$
$cm^{-3}$) \cite{jacoboni1977review}. It is worth to note that the FLA samples have a larger substitutional fraction and a smaller carrier concentration than that of PLA samples. This discrepancy might be due to the fact that in PLA samples selenium diffuses out- and in-ward resulting in a larger effective doping depth. On the other hand the interstitial selenium or other defects probably also contribute to free electrons as for the case of titanium supersaturated Si \cite{sanchez2009assessment,pastor2012interstitial}.

\begin{figure}
\includegraphics[scale=0.4]{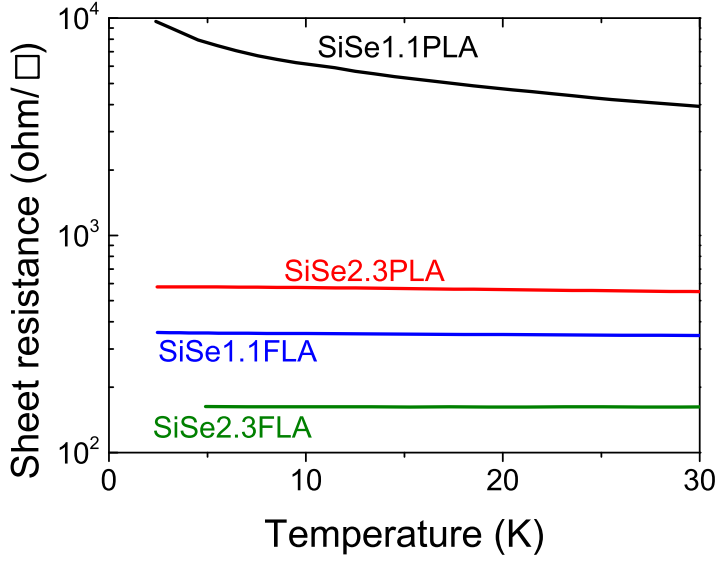}
\caption{Temperature dependent sheet resistance of selenium implanted Si
annealed by FLA (1.3 ms, 3.4 kV) or PLA (308 nm, 28 ns, 0.9 J/cm$^2$): With increasing selenium concentration, an
insulator-metal transition occurs for the PLA samples, while all FLA samples show quasi-metallic conduction.}\label{fig_RvsT}
\end{figure}

\begin{figure}
\includegraphics[scale=0.4]{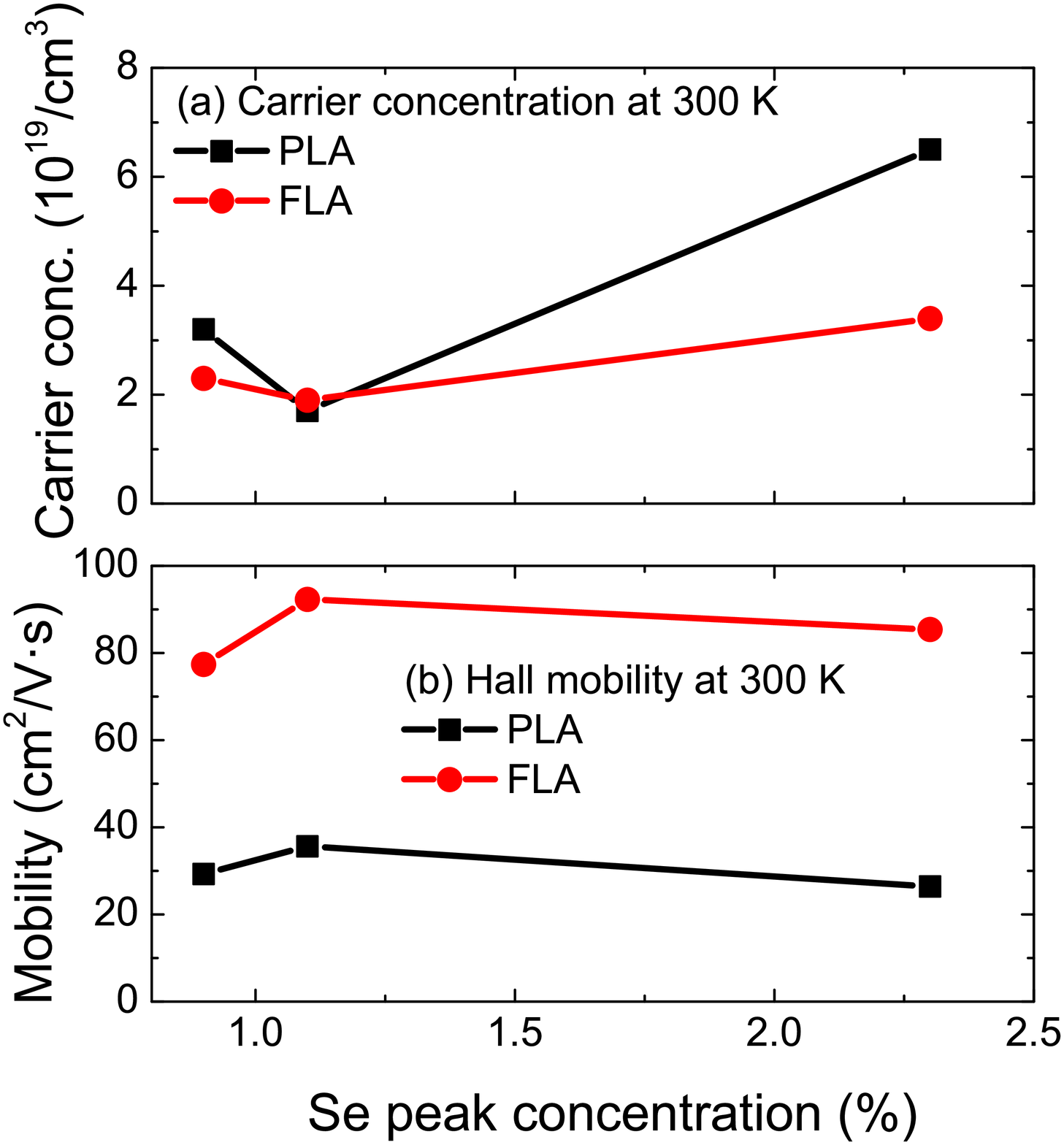}
\caption{(a) The carrier concentration and (b) the Hall mobility at 300 K of selenium implanted Si
annealed by FLA (1.3 ms, 3.4 kV) or PLA (308 nm, 28 ns, 0.9 J/cm$^2$) with optimized parameters. The mobility of
FLA samples is comparable with the shallow donor considering the
dopant concentration
\cite{jacoboni1977review}. The solid lines are only for guiding the eyes.}\label{fig_mobility300K}
\end{figure}

%It is worthy to note the difference between samples SiSe0.9 and
%SiSe1.1. SiSe0.9 has a smaller Se concentration but a larger
%effective doping thickness than SiSe1.1. However, SiSe0.9 (not
%shown in Fig. \ref{fig_RvsT}), after either FLA or PLA, exhibits a
%larger conductivity, a similar or larger carrier concentration
%than sample SiSe1.1. It indicates that a thicker implantation
%layer favors a high doping efficiency. This also explains the
%slight discrepancy in our data compared with Ref.
%\onlinecite{PhysRevLett.108.026401}, in which the implantation
%depth is around 350 nm.

\section*{Discussion}

How to understand the success of hyperdoping Si with selenium via
solid phase epitaxy? Let us revisit the mechanism for realizing
hyperdoped semiconductors. The hyperdoping is attributed to the
so-called solute trapping at the moving amorphous/crystalline
interface when the dopant retain time is larger than the time
required for one monolayer regrowth
\cite{CampisanoAPL1980,duffy2005impurity,wundisch2009millisecond}.
This process is determined by the diffusion coefficient, which is
much larger in liquid phase than in solid phase for most of
dopants. This reminds us that a compromise approach might exist
for chalcogen elements which are relatively slow diffusers in the
solid phase compared with transition metals. Indeed,
supersaturation of tellurium in Si up to 3$\times$10$^{20}$
cm$^{-3}$ has been realized by furnace annealing at 550
$^{\circ}$C and the substitutional fraction is around 70\% in
tellurium implanted Si \cite{CampisanoAPL1980}. However, the
substitutional fraction is largely decreased to 45\% when the
implantation fluence is increased as for the case of selenium
implanted Si \cite{0256-307X-29-9-097101}.

We try to characterize the competition between the solute trapping
and diffusion by estimating the time needed to regrow ($\tau_G$)
or to diffuse ($\tau_D$) over a Si monolayer (0.27 nm). In other
words, the speed of the resolidification and the speed at which
the impurity atoms can move determine how likely they will stay
ahead or be trapped by the moving amorphous/crystalline interface.
If $\tau_D$ is larger than $\tau_G$, the dopants are able to be
trapped in the crystalline matrix. In Figure 6, we compare $\tau_G$ and $\tau_D$
estimated from data published in literature. $\tau_D$ is
calculated according to the data in ref.
\cite{Vydyanath1978}. The large uncertainty in $\tau_G$
comes from the large scattering in the regrowth velocity, which
exhibits different values reported by various groups
\cite{Kokorowski_JAP1982,lietoila1981rate,Roth_APL1990,csepregi2008substrate,CampisanoAPL1980}.
However, as shown in Fig. 6, in
solid phase Si $\tau_D$ is generally larger than $\tau_G$. That
means selenium impurities can be trapped in the Si matrix if an
optimized thermal treatment is applied even in solid phase
processing. Particularly, in the low temperature regime, $\tau_D$
is much larger than $\tau_G$, which well explains the realization
of doping above solubility limit by low temperature annealing
\cite{CampisanoAPL1980}.

\begin{figure}
\includegraphics[scale=0.4]{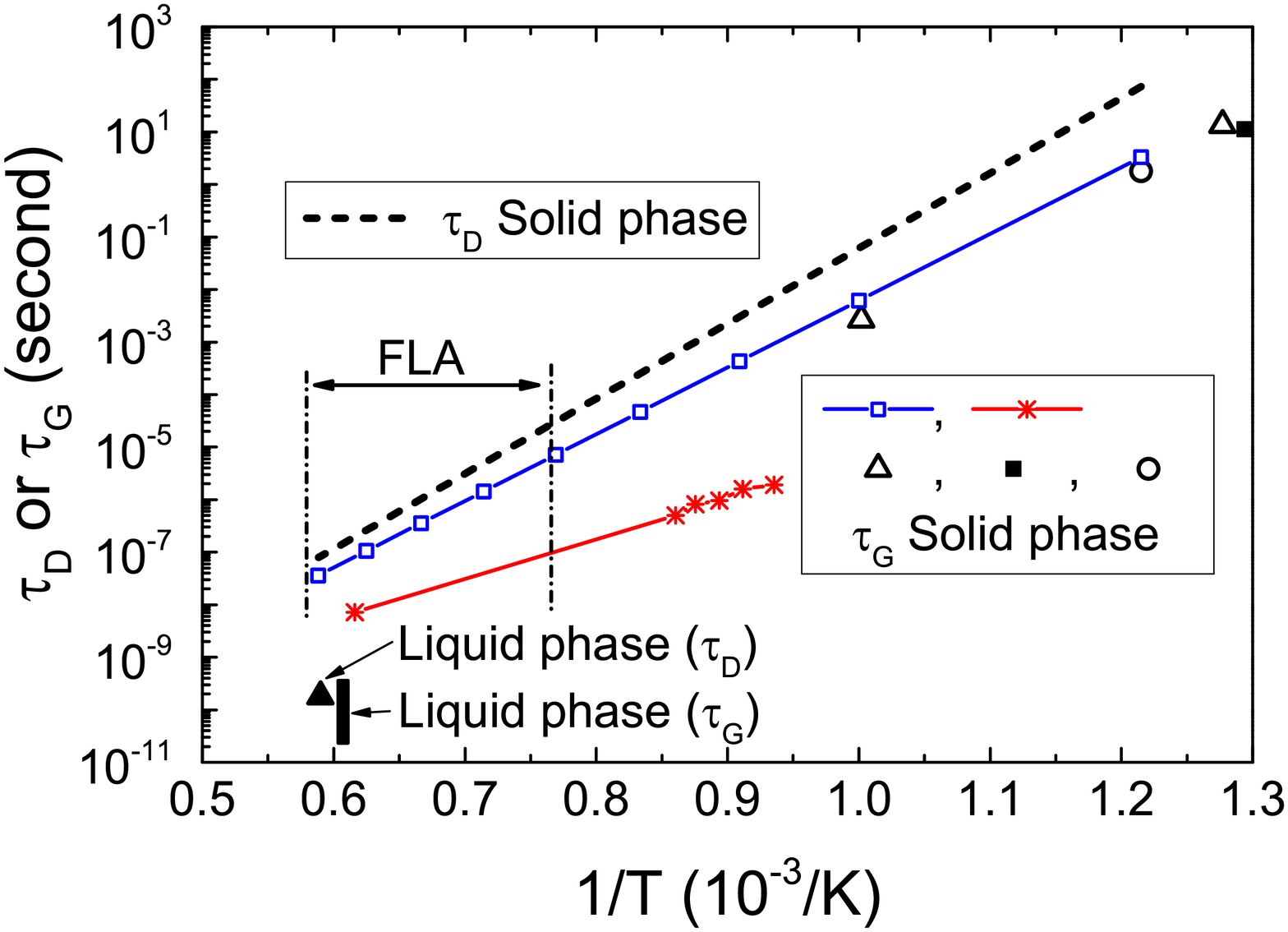}
\caption{Competition between the Si recrystallization and selenium
diffusion characterized by the time needed to regrow ($\tau_G$) or
diffuse ($\tau_D$) one monolayer. The dashed line is $\tau_D$ in
solid phase according to the diffusion parameter in Ref \cite{Vydyanath1978}. The other
lines or symbols are $\tau_G$ according to different references: open square \cite{Kokorowski_JAP1982}, star \cite{lietoila1981rate}, open triangle \cite{Roth_APL1990}, solid square \cite{csepregi2008substrate}, open circle  \cite{CampisanoAPL1980}.
Despite the possible uncertainty of the regrowth velocity,
$\tau_G$ is smaller than $\tau_D$. That means it is possible to
trap selenium and realize metastable, selenium over-saturated Si layer. The vertical dashed and dotted lines indicate the working regime of flash lamp annealing (FLA). The reported $\tau_G$ is shown as a solid triangle \cite{BobJAP2010} and $\tau_D$ is shown as a vertical thick line by assuming the growth velocity of 1--10 m/s for pulsed laser annealing (PLA).
}\label{fig_growth_diffusive_time}
\end{figure}

Another criterion to be considered is the annealing duration. In
the regrown metastable layer, the impurity concentration is much
above the thermal equilibrium solubility limit. During the
prolonged annealing to finish the regrowth completely, the
metastable solubility returns to the equilibrium value as dopants
come out of their substitutional positions. This has been observed
in sulfur hyperdoped Si \cite{Simmons128741}. The decrease in the
substitutional fraction with increasing dopant concentration is
due to incipient precipitation \cite{Campisano1980BiSi}. For a
first approximation, the metastable phase does not precipitate if
the mean diffusion length of the impurity at the annealing
temperature is less than their average distance. For a diffusion
length larger than the average impurity distance, there is a
finite probability of nucleation of a secondary phase or dimers.
Therefore, the diffusion length $L=(2Dt)^{1/2}$, D being the
diffusion coefficient \cite{Vydyanath1978} and $t$ the annealing time, must be relatively short in order to achieve a high concentration of dopants on the substitutional sites. We
illustrate the estimated results in Fig. 7, in which the working regimes of furnace
annealing (FA), rapid thermal annealing (RTA) and flash lamp
annealing (FLA) are indicated. In the case of FA at low
temperature (around 500 $^{\circ}$C), an annealing duration more
than 1000 s is needed to regrow the whole layer. A longer
annealing time and a slightly higher annealing temperature
strongly increase the diffusion length. From Fig. 7, an annealing processing in millisecond
or microsecond while keeping the system in solid phase could
result in a minimized diffusion length. This also explains why
the substitutional fraction strongly reduces in furnace annealed
selenium-implanted samples when the selenium fluence is increased
\cite{0256-307X-29-9-097101}.

\begin{figure}
\includegraphics[scale=0.4]{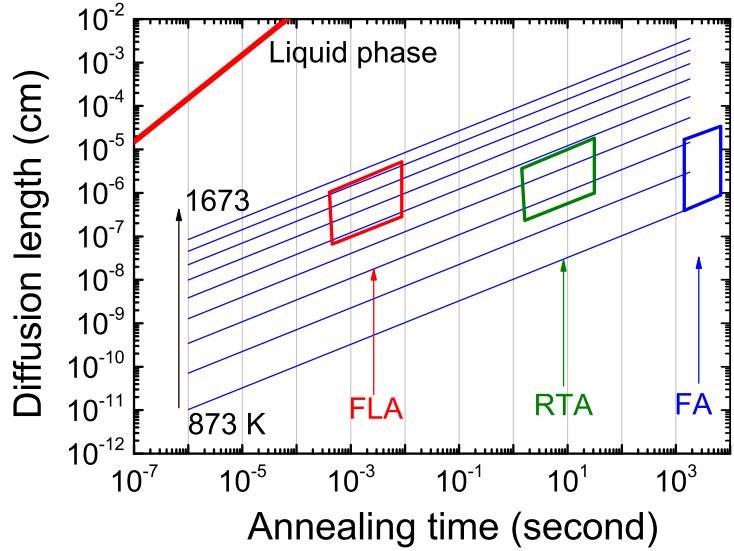}
\caption{Selenium diffusion length ($L$) in Si at different temperature
vs. time duration. The three boxes indicate the working regime of
different thermal process in solid phase: low temperature furnace
annealing (FA) for more than 1000 second, rapid thermal annealing (RTA) for seconds and flash lamp annealing (FLA) for milliseconds at high temperature. The diffusion length of selenium in liquid phase is also shown for comparison.}\label{fig_diffusionlength}
\end{figure}

In summary, we have presented a solid phase epitaxy approach by millisecond
FLA to realize selenium hyperdoped Si from implanted amorphous layers.
The dopant redistribution, which always occurs in pulsed-laser
induced liquid phase epitaxy, can be effectively suppressed by
FLA. The FLA-prepared sample exhibits larger Hall mobility and
conductivity than the samples prepared by liquid phase epitaxy.
The success of FLA lies in the facts that the selenium impurities
have moderate diffusion coefficients in solid Si and that FLA
occurs within milliseconds (much shorter than furnace or rapid
thermal annealing). The regrowth velocity can beat the
dopant diffusion in carefully optimized annealing condition. Our
finding is not limited to selenium, but generally
interesting for chalcogen impurities and other unconventional dopants in semiconductors.
Annealing in the microsecond or millisecond range might be
optimized for realizing hyperdoping of transition metal impurities
in Si as well as in III-V compound semiconductors.

\section*{Methods}

Semi-insulating (100) Si wafers were implanted with selenium at room
temperature. The implantation energy and fluence are listed in
table I. The resistivity of the virgin Si wafer is above 4000 $\Omega$$\cdot$cm at room temperature. This resistivity corresponds to a sheet resistance of 8$\times$10$^4$ $\Omega/square$ at room temperature, which is even 10 times larger than sample SiSe1.1PLA (shown in Fig. 4) with the most resistive sample in our experiment. Therefore, the parallel conductivity from the Si substrate can be neglected.

Implanted Si samples were flash-lamp annealed for 1.3, 3 and 20 ms. The FLA system employed in our experiments has been
introduced in details in Ref. \cite{skorupa2005advanced,mcmahon2007flash}. It consists of two annealing systems. At the top there are twelve 30 cm long xenon (Xe) lamps spaced by about 3 cm representing together with the reflector the FLA system, and at the bottom a lower bank of halogen lamps allows the wafer to be preheated to a selected temperature - a type of rapid thermal annealing system. The Xe lamps are energized by discharging a capacitor/inductor unit in the millisecond time scale. The maximum energy which can be delivered to the sample during a single 20 ms shot is 250 kJ which is sufficient to melt silicon. The emission spectrum of Xe lamps used in our system falls in the visible spectral range in between 350 and 800 nm. For effective annealing, FLA treated materials should have a high absorption coefficient in this spectral region, e.g. silicon is perfectly suited for FLA processes. Wafers up to 100 mm in diameter can be processed with a lateral homogeneity better than 5\%. The temperature simulation was carried out based on the wave transfer matrix method for modelling the absorption of the flash light, and the numerical solution of the one-dimensional heat equation \cite{smith2004modeling}. The optical system taken into consideration for temperature simulation comprises the inert gas atmosphere of argon, amorphous Si layer formed during ion implantation and the bulk Si. The energy delivered to the sample during the FLA process depends on the overall absorption and transmission of the investigated system. The as-implanted Si absorb about 85\% of the incident flash lamp spectrum, while the virgin Si wafer absorb only about 65\% of the light delivered to the sample surface. Therefore the annealing parameters used to obtain the same final temperature during the FLA process in different materials have to be carefully adjusted. As one example, according to the simulation of the temperature distribution in Si wafer during 3 ms annealing, the maximum temperature at the surface is reached after 2.5 ms with temperature gradient of about 1 $^{\circ}$C/$\mu$m.

The laser annealing system is a Coherent XeCl excimer laser with
308 nm wavelength and 28 ns duration which was focused to a 5 mm
$\times$5 mm square spot, where the lateral intensity variation
was less than ±3\%. Samples were irradiated with different energy
fluences ranging from 0.5 J/cm$^{2}$ to 1.2 J/cm$^{2}$.

The structural properties of the annealed samples were
investigated by Raman spectroscopy, Rutherford backscattering
spectrometry/channeling (RBS/channeling) and X-ray diffraction
(XRD). The RBS measurements were performed with a collimated 1.7
MeV He$^+$ beam at a backscattering angle of 170$^{\circ}$. For the results in the inset of Fig. 1(b), the measurements were performed with 1.5 MeV He$^+$ ions. The
sample was mounted on a three-axis goniometer with a precision of
0.01$^{\circ}$. The channeling spectra were measured by aligning
the sample to make the impinging He$^+$ beam parallel with the
Si[001] axis. The XRD experiment were performed at Rossendorf
Beamline (ROBL), European Synchrotron Radiation Facility (ESRF),
using a X-ray with wavelength of 0.1078 nm. The phonon modes were
determined by Raman spectroscopy in a backscattering geometry in
the range of 200 to 600 cm$^{-1}$ using a 532 nm Nd:YAG laser with
a liquid nitrogen cooled charge coupled device camera.
Electrical properties were measured in van der Pauw geometry
using a commercial Lakeshore Hall System. Gold electrodes were sputtered onto the four corners of the square-like samples. The sputtering process also removed the nature SiO$_2$ layer to some extent. Silver glue was used to contact the wires to the gold electrodes. All contacts are confirmed to be ohmic as we checked by measuring current-voltage curves at different temperatures.

%\bibliography{../SeSi_liquid_vs_solid}

\section*{Acknowledgements}
We would like to thank the ion implanter group at HZDR for
technical support and Olga Roshchupkina for her assistance with
synchrotron XRD. The Helmholtz-Association (HGF-VH-NG-713) is
gratefully acknowledged. The author (FL) thanks the support by
China Scholarship Council (File No. 201307040037). The discussion
with Karl-Heinz Heinig is greatly acknowledged.

\section*{Author Contributions Statement}
SZ designed the research and wrote the manuscript. FL did the pulsed laser annealing and RBS measurements. SP performed flashlamp annealing. KG and MK carried out the Raman and transport measurements, respectively. CB contributed to the XRD measurement and data analysis. MP contributed to the thermal dynamic estimation. WS and MH supervised the work. All authors contributed to discussion and correction for the manuscript.

The authors declare no competing financial interests.

Correspondence and requests for materials
should be addressed to Shengqiang Zhou~(email: s.zhou@hzdr.de)

\begin{table*}
\caption{\label{tab:sample} Sample definition and related
parameters. The samples are referred as SiSe0.9PLA annealed by
pulsed laser or SiSe0.9FLA annealed by flash lamp, respectively,
with corresponding optimal parameters. The depth distribution of
selenium (estimated thickness) is calculated using SRIM and verified by
RBS measurements.}
%\begin{ruledtabular}

\begin{tabular}{cccc}
  \hline\hline
  Sample ID  & Implantation parameters &  Estimated thickness & Selenium peak concentration (\%) \\
  \hline
  SiSe0.9 & 110 keV, 2.8$\times$10$^{15}$ cm$^{-2}$ & $\sim$150 nm & 0.9 \\
  & 50 keV, 1.4$\times$10$^{15}$ cm$^{-2}$&  &  \\
  \hline
  SiSe1.1 & 60 keV, 2.5$\times$10$^{15}$ cm$^{-2}$& $\sim$100 nm & 1.1  \\
 \hline
  SiSe2.3 & 60 keV, 5.0$\times$10$^{15}$ cm$^{-2}$& $\sim$100 nm & 2.3 \\
 %SeSi4 & 60 keV 1.2E16 & 50 nm & 0.056  \\
 \hline\hline
\end{tabular}
%\end{ruledtabular}
\end{table*}

\end{document}